\documentclass[amsmath,amssymb,aps,letterpaper,prl,twocolumn,longbibliography,10pt]{revtex4-1}
\pdfoutput=1
\usepackage{graphicx}
\usepackage{color}
\usepackage{dsfont}
\usepackage{simplewick}

\usepackage{subfigure}

\usepackage{color}
\usepackage{enumitem}

\makeatletter \let\@keywords\@empty \let\@subject\@empty
\providecommand{\keywords}[1]{\gdef\@keywords{#1}}
\providecommand{\subject}[1]{\gdef\@subject{#1}}
\def\thetitle{\@title}
\def\theauthor{\@author}
\def\thesubject{\@subject}
\def\thedate{\@date}
\def\thekeywords{\@keywords}
\makeatother
\AtBeginDocument{%
\hypersetup{pdftitle={\thetitle}}%
\hypersetup{pdfauthor={\theauthor}}%
\hypersetup{pdfsubject={\thesubject}}%
\hypersetup{pdfkeywords={\thekeywords}}%
}

\usepackage[bookmarks=true,hyperfigures=true]{hyperref}
\usepackage{fixmath}
\usepackage{mathtools}

\providecommand{\href}[2]{#2}

\let\oldbfseries=\bfseries
\let\oldmdseries=\mdseries
\let\oldnormalfont=\normalfont
\renewcommand{\bfseries}{\oldbfseries\boldmath}
\renewcommand{\mdseries}{\oldmdseries\unboldmath}
\renewcommand{\normalfont}{\oldnormalfont\unboldmath}

\makeatletter
\newlength{\apb@width}
\newcommand{\autoparbox}[2][c]{\settowidth{\apb@width}{#2}\parbox[#1]{\apb@width}{#2}}

\makeatother

\RequirePackage{verbatim}

\makeatletter
\newwrite\bibinl@out
{\immediate\closeout\bibinl@out\@esphack}
\makeatother

\newcommand{\de}{\operatorname{d}\!}
\DeclareMathOperator{\tr}{tr}

\newcommand{\ihalf}{\frac{i}{2}}

\renewcommand{\digamma}{\Psi}

\newcommand{\eqndot}{\, . }
\newcommand{\eqncom}{\, , }

\newcommand{\YM}{{\mathrm{\scriptscriptstyle YM}}}

\DeclareMathOperator{\phaneq}{\phantom{{}=}}
\DeclareMathOperator{\idm}{\mathds{1}}

\begin{document}

\title{Asymptotic one-point functions in AdS/dCFT}

 \author{Isak Buhl-Mortensen}%
  \email{buhlmort@nbi.ku.dk}
    \author{Marius de Leeuw}%
     \email{deleeuwm@nbi.ku.dk}
        \author{Asger C.\ Ipsen}%
  \email{asgercro@nbi.ku.dk}
       \author{Charlotte Kristjansen}
         \email{kristjan@nbi.ku.dk}
 \author{Matthias Wilhelm}%
   \email{matthias.wilhelm@nbi.ku.dk}

\affiliation{%
Niels Bohr Institute, Copenhagen University,\\
Blegdamsvej 17, 2100 Copenhagen \O{}, Denmark
}%

\begin{abstract}
We take the first step in extending the integrability approach to one-point functions in AdS/dCFT to higher loop orders. More precisely,
we argue that the formula encoding all tree-level one-point functions of SU(2) operators in  the defect version of  ${\cal N}=4$ SYM theory,  dual to the D5-D3 probe-brane system with flux, has a natural asymptotic generalization to higher loop orders.
 The asymptotic formula correctly encodes the information about the one-loop correction to the one-point functions of non-protected operators once dressed by a simple flux-dependent factor, as we demonstrate by an explicit computation involving a novel object denoted as an amputated matrix product state. Furthermore, when applied to the BMN vacuum state, the asymptotic formula gives a result for the one-point function which in a certain double-scaling limit agrees with that obtained in  the dual string theory up to wrapping order. 
 \end{abstract}

\maketitle

\section{Introduction}

Apart from observables which are protected by supersymmetry, the AdS/CFT correspondence has not provided us with
many examples of
quantities which can be explicitly calculated to all orders in the coupling constant in both string theory and  field
theory and successfully matched. The main examples are the cusp anomalous dimension~\cite{Freyhult:2010kc} and  the
expectation value of the circular Maldacena-Wilson loop~\cite{Drukker:2000rr,Drukker:2000ep,Semenoff:2001xp}.  An instructive attempt to arrange for a situation which could
allow an all-order comparison between gauge and string theory was made with the invention of the Berenstein-Maldacena-Nastase (BMN) limit, where a certain double-scaling parameter combining the 't Hooft coupling constant $\lambda$ with a large angular momentum quantum number
was introduced and certain observables being close to protected were considered~\cite{Berenstein:2002jq}. 
However, it turned out that for the observables considered the BMN expansion became inconsistent starting at four-loop order in the field theory~\cite{Bern:2006ew,Beisert:2006ez,Cachazo:2006az}. 

In a variant of the AdS/CFT correspondence which involves
a D5-D3 probe-brane set-up on the string-theory side and a co-dimension-one defect in ${\cal N}=4$ supersymmetric Yang-Mills (SYM) theory, another double-scaling limit
has recently been proposed~\cite{Nagasaki:2011ue}. It consists of sending the 't Hooft coupling as well as a certain 
background gauge field flux $k$ to  infinity while keeping a certain ratio involving the two parameters fixed. While the study of 
the BMN expansion acted as a seed for the development of the integrability approach to ${\cal N}=4$ SYM theory~\cite{Beisert:2010jr}, at
the present stage we already have available a vast amount of integrability tools that we can make use of when investigating
the defect set-up and the associated novel double-scaling limit. In addition, in the defect case we have an entirely new collection of observables including one-point functions, two-point functions between operators of unequal conformal dimension and 
correlators between bulk and boundary fields~\cite{Cardy:1984bb}. In particular, we can consider the BMN vacuum states,
 BPS states of ${\cal N}=4$ SYM theory whose two- and three-point functions do not get quantum corrections
in pure ${\cal N}=4$ SYM theory but whose one-point functions are non-vanishing and receive quantum corrections in the defect theory.

One-point functions of protected operators were calculated at tree level in the above-mentioned defect CFT in~\cite{Nagasaki:2012re} and in a closely related theory building on a non-supersymmetric D7-D3 probe-brane system in~\cite{Kristjansen:2012tn}.
Furthermore, exploiting the integrability structure of
${\cal N}=4$ SYM theory and introducing an appropriate boundary state in the form of a matrix product state,
one-point functions of non-protected operators were calculated at tree level for the SU(2) sector in~\cite{deLeeuw:2015hxa,Buhl-Mortensen:2015gfd}.  This approach was generalized to the SU(3) sector \cite{deLeeuw:2016umh} as well as to the SO(6) sector of the above-mentioned non-supersymmetric defect CFT~\cite{deLeeuw:2016ofj}.%
\footnote{Note that the latter two sectors are only closed at one-loop level.}
Most recently, the one-loop correction to the
one-point function of the BMN vacuum was calculated~\cite{Buhl-Mortensen:2016pxs,Buhl-Mortensen:2016jqo} 
and shown to match the string-theory prediction of~\cite{Nagasaki:2012re}. 
In addition,
a strategy for computing the one-loop correction to the one-point functions of non-protected operators was presented~\cite{Buhl-Mortensen:2016jqo}.
This involved the introduction of a new object denoted as the amputated matrix product state.

In the present letter, we will argue that the integrability approach to one-point functions suggests a certain  
generalization of the tree-level formula for the SU(2) sector to higher loop orders. 
We shall furthermore concretely implement the above-mentioned strategy for the calculation of one-loop corrections to one-point functions and show that the results can be accounted for by the suggested asymptotic formula when dressed by a simple
flux-dependent factor.
This flux factor leads to a breakdown of the above-mentioned double-scaling limit for non-protected operators already at one-loop order.
For protected operators, the flux factor is absent and we will show that the proposed formula implies that the 
one-point function of the BMN vacuum state has an expansion 
that in the double-scaling limit, up to terms of wrapping order, matches an expansion derived in the string-theory language using a supergravity approximation.

\section{Our proposal}

The defect version of ${\cal N}=4$ SYM theory which is dual to the D5-D3 probe-brane system with flux $k$ is characterized by
having a co-dimension-one defect, say at $x_3=0$, separating two regions of space, $x_3>0$ and $x_3<0$, where
the gauge group is respectively (broken) U($N$) and U($N-k$).  The difference in the rank of the gauge group implies assigning the
following vacuum expectation values, for $x_3>0$, to three out of the six scalar fields of ${\cal N}=4$ SYM theory:
\begin{align}
\langle \phi_i\rangle_{\text{tree}} = -\frac{1}{x_3} t_i \oplus 0_{(N-k)\times (N-k)}, \hspace{0.5cm} i=1,2,3,
\end{align}
where the $t_i$ are the generators of a $k$-dimensional irreducible representation of SU(2).
For a precise description of the holographic set-up, we refer to~\cite{Buhl-Mortensen:2016jqo}
as well as the original papers~\cite{Constable:1999ac,Karch:2000gx}. 

As usual, we identify two complex scalars of $\mathcal{N}=4$ SYM theory with spins of an integrable SU(2) spin chain as $\uparrow\,\equiv X=\phi_1+i\phi_4$ and $\downarrow\,\equiv Y=\phi_2+i\phi_5$.
A Bethe (eigen)state of this spin chain is characterized by two Dynkin labels $L,M$ corresponding respectively to the length and the number of excitations, and in addition by $M$ rapidities $\{u_i\}$ that satisfy certain Bethe equations.
For a given eigenstate $|\mathbf{u}\rangle$, we define the corresponding single-trace operator from the SU(2) sector as
\begin{equation}
  \mathcal{O} \equiv  \left(\frac{4\pi^2}{\lambda}\right)^{\frac{L}{2}}\frac{\mathcal{Z}}{\sqrt{L}}
    \frac{\tr\prod_{l=1}^L\Big(\langle\uparrow_l \!\! |\otimes X+\langle\downarrow_l \!\! |\otimes Y\Big) | \mathbf{u} \rangle }{\sqrt{ \langle \mathbf{u} | \mathbf{u} \rangle}}\eqndot
\end{equation}
Far away from the defect, the tree-level two-point function of $\mathcal{O}$ is normalized to unity, 
and we will use the freedom in the
choice of the finite part of the renormalization constant $\mathcal{Z}$ to enforce this also at loop level. 
The one-point function then takes the form
\begin{equation}
  \langle \mathcal{O}(x) \rangle = \left(\frac{4\pi^2}{\lambda}\right)^{\frac{L}{2}}\frac{C_k}{\sqrt{L}}  \frac{1}{x_3^\Delta}\eqncom
\end{equation}
where $\Delta$ denotes the scaling dimension of the operator. The calculation of $C_k$ will be the subject of this letter.

\paragraph{Tree level} 
At tree level, the one-point function can be written as the overlap of a Bethe eigenstate of the Heisenberg spin chain with a matrix product state \cite{deLeeuw:2015hxa,Buhl-Mortensen:2015gfd}.
The corresponding Bethe equations read
\begin{align}
1 =\left(\frac{u_k-\frac{i}{2}}{u_k+\frac{i}{2}}\right)^L \prod_{\substack{j=1\\j\neq k}}^M \frac{u_k-u_j+i}{u_k-u_j-i} \equiv \exp[ i \Phi_k]\eqndot
\end{align}
Using the algebraic Bethe ansatz approach \cite{Faddeev:1996iy}, the Bethe state can be built from the ferromagnetic vacuum $|0\rangle_L$ with all spins up via the creation operators $B(u)$:
\begin{equation}
 |\mathbf{u}\rangle=B(u_1)\cdots B(u_M)|0\rangle_L\eqndot
\end{equation}
Defining the matrix product state as
\begin{equation}
\label{MPS}
 \langle \text{MPS}|=\tr\prod_{l=1}^L\Big(\langle\uparrow_l \!\! |\otimes t_1+\langle\downarrow_l \!\! |\otimes t_2\Big)\eqncom
\end{equation}
the tree-level one-point function of $\mathcal{O}$ is  given as
\begin{align}
 C_k= \frac{\langle \text{MPS}|\mathbf{u}\rangle}{\sqrt{\langle\mathbf{u}|\mathbf{u}\rangle}}\eqndot
 \label{Ck}
\end{align}

In \cite{deLeeuw:2015hxa}, it was shown that only operators with $L$ and $M$ even and with paired rapidities $\{u_i\} = \{-u_i\}$ have non-trivial one-point functions \footnote{For this reason, we have left out a factor $(-1)^L$ in \eqref{Ck}.}.
For $k=2$, the tree-level one-point function can be elegantly described in terms of the Bethe function $\Phi$ introduced above. Let us order the roots as $\{u_1,\ldots, u_{\frac{M}{2}}, -u_1,\ldots, -u_{\frac{M}{2}}\}$ and introduce the following $\frac{M}{2}\times\frac{M}{2}$ dimensional matrices $G_{\pm}$:
\begin{align}
&G_{\pm} = \partial_m \Phi_n \pm \partial_{m+\frac{M}{2}} \Phi_n\eqncom
\end{align}
with $\partial_m \equiv \frac{\partial }{\partial u_m}$. Then, the one-point function for $k=2$ can be written as
\begin{align}\label{eq:SU2quotient}
C_2 = 2^{1-L}\sqrt{ \frac{Q(\ihalf)}{Q(0)}}\sqrt{ \frac{\det  G_+}{\det  G_-}} \eqncom
\end{align}
where  $Q(u) = \prod_{i=1}^M (u-u_i)$ is the Baxter polynomial. 

According to \cite{Buhl-Mortensen:2015gfd}, the one-point function for $k>2$  then takes the form
\begin{align}\label{eq:SU2genK}
C_k =i^L T_{k-1}(0)\sqrt{\frac{Q(\ihalf)Q(0)}{Q^2(\frac{ik}{2})} }\sqrt{\frac{\det  G_+}{\det  G_-}} \eqncom
\end{align}
where 
\begin{align}\label{eq:Transfer}
T_n(u) =
\!\! \sum_{a=-\frac{n}{2}}^{\frac{n}{2}}\!\! (u+ia)^L\frac{Q(u+\frac{n+1}{2}i)Q(u-\frac{n+1}{2}i)}{Q(u+(a-\frac{1}{2})i)Q(u+(a+\frac{1}{2})i)}
\end{align}
can be identified as the transfer matrix of the Heisenberg spin chain in the $(n+1)$-dimensional representation
\footnote{K. Zarembo,  Defect CFT and Integrability, talk at ``New Trends in Integrable Models'', Natal, Brazil, Oct.\ 2016.}.

\paragraph{Quantization} 
Bearing in mind the integrability approach to the spectral problem of ${\cal N}=4$ SYM theory, it is natural to introduce the coupling constant dependence via the Zhukovsky variable $x$~\cite{Beisert:2005fw}:
\begin{align}
&x + \frac{1}{x} = \frac{u}{g},
&&x = \frac{u}{g} - \frac{g}{u} + O(g^2)\eqncom
\end{align}
where the effective planar coupling constant $g^2$ is related to the 't Hooft coupling $\lambda=Ng_\YM^2$ as $g^2=\frac{\lambda}{16\pi^2}$ and where the cut of the function $x(u)$ is taken to be the straight line $[-2g,2g]$.
The all-loop asymptotic Bethe equations which determine the conformal operators of ${\cal N}=4$ SYM theory and their anomalous dimensions
are then given by \cite{Beisert:2006ez}:
\begin{align}\label{eq:BAEquantum}
1 = &\left(\frac{x(u_k-\frac{i}{2})}{x(u_k+\frac{i}{2})}\right)^L \prod_{j\neq k} \frac{u_k-u_j+i}{u_k-u_j-i} \, \nonumber 
\exp (2i \theta(u_k,u_j)) \\
 \equiv &\exp [i\tilde{\Phi}_k] \eqncom
\end{align}
where $\exp(2i\theta(u_k,u_j))$ is the so-called dressing phase.
A natural generalization of \eqref{eq:SU2quotient} is  obtained by replacing the 
classical Bethe function $\Phi$ by the quantum Bethe function $\tilde{\Phi}$. Furthermore, 
a natural generalization of the the transfer matrix is the following one
\begin{multline}\label{eq:Transfer quantum}
\tilde{T}_n(u) = g^L \sum_{a=-\frac{n}{2}}^{\frac{n}{2}} x(u+ia)^L\\\times\frac{Q(u+\frac{n+1}{2}i)Q(u-\frac{n+1}{2}i)}{Q(u+(a-\frac{1}{2})i)Q(u+(a+\frac{1}{2})i)}\eqndot
\end{multline}
This gives a natural expression for  \eqref{eq:SU2genK} at the quantum level. Of course, the roots $u_i$ appearing in the Baxter polynomials satisfy the all-loop Bethe equations \eqref{eq:BAEquantum}.  It is  not excluded that further modifications
of the transfer matrix are necessary but for the consistency checks that we perform the present modification suffices.  Furthermore, 
the phase factor in (\ref{eq:BAEquantum}) does not come into play in these checks.
However,  we do need to allow for a flux factor $\mathbb{F}_k$, such that we find
\begin{align}\label{eq:Ansatz}
C_k =  i^L\tilde{T}_{k-1}(0)\sqrt{\frac{Q(\ihalf)Q(0)}{Q^2(\frac{ik}{2})} }\sqrt{\frac{\det  \tilde{G}_+}{\det  \tilde{G}_-}} 
\,\mathbb{F}_k\eqndot
\end{align}
The flux factor $\mathbb{F}_k$ is $1$ for protected operators and its general form at one-loop order turns out to be
\begin{align}
\mathbb{F}_k = 1+g^2 \Big[ \Psi(\textstyle{\frac{k+1}{2}}) + \gamma_E - \log 2 \Big] \Delta^{(1)}+O(g^4)\,,
\end{align}
where $\Delta^{(1)} = 2\sum_{i=1}^M\frac{1}{u_i^2+\frac{1}{4}}$ is the one-loop correction to the scaling dimension.
Note that the Euler digamma function $\Psi$ can be reexpressed in terms of the harmonic number $H$, which is generalized to non-integer   arguments via $H_x=\Psi(x+1)+\gamma_E$. 

\section{Checks}
We now test (\ref{eq:Ansatz}) -- first at one-loop order for non-protected operators in the SU(2) sector and then at higher orders for the BMN vacuum. Finally, we will discuss the flux factor and the fate of the double-scaling limit.

\subsection{SU(2) at one-loop}
\label{subsec: SU(2) one-loop}

In \cite{Buhl-Mortensen:2016jqo}, we have shown that the one-loop one-point function is given by the sum of three contributions: a) the manifestly finite overlap of the Bethe eigenstate with a special spin-chain state, denoted
as an amputated matrix product state, b) an ultraviolet (UV)-divergent contribution proportional to the one-loop dilatation operator, which requires operator renormalization, and c) the one-loop correction to the Bethe state.
Demanding that the two-point function far away from the defect remains unit-normalized also at one-loop order
fixes the renormalization constant to be
$\mathcal{Z}=1+g^2\frac{\Delta^{(1)}}{2}\left(\frac{1}{\epsilon}+1+\gamma_E+\log\pi\right)+O(g^4)$, see for instance \cite{Beisert:2003tq}. 
The one-loop one-point function then reads \cite{Buhl-Mortensen:2016jqo}
\begin{align}\label{eq:one-loop}
C_k &=\frac{(\langle \text{MPS}|+ g^2\langle \text{AMPS}|)|\mathbf{u}\rangle}{\sqrt{\langle\mathbf{u}|\mathbf{u}\rangle}}\times\\
&\phaneq\Big[1+g^2\left(\Psi({\textstyle\frac{k+1}{2}})+\gamma_E-\log 2+{\textstyle\frac{1}{2}}\right)\Delta^{(1)}\Big]
+ O(g^4)
\eqncom
\nonumber
\end{align}
where $|\text{AMPS}\rangle$ denotes the amputated matrix product state, to be explicated below, and $|\mathbf{u}\rangle$ denotes the loop-corrected Bethe state.
In order to evaluate \eqref{eq:one-loop} explicitly, we need two ingredients. We need to evaluate the overlap of $\langle \text{AMPS}|$ with the Bethe state and we need to compute the first correction to the Bethe state,
 i.e.\ the two-loop Bethe eigenstate.

\paragraph{Overlap with $\langle\mathrm{AMPS}|$} The amputated matrix product state $\langle \text{AMPS}|$ is defined as~\cite{Buhl-Mortensen:2016jqo}
\begin{equation}
 \langle \text{AMPS}|= \sum_{l=1}^L\mathcal{A}_{l,l+1}\langle \text{MPS}|\eqncom \label{AMPS}
\end{equation}
where $\mathcal{A}_{i,i+1}$ removes the matrices at positions $i$ and $i+1$ (with $L+1\sim1$) if they are identical and otherwise kills the trace, cf.\ \eqref{MPS}.

Let us consider the overlap between a Bethe state and the 
amputated matrix product state. The overlap is only non-zero for an even number of magnons $M$, and in the coordinate formulation it reads
\begin{equation}
\begin{aligned}
	\langle \text{AMPS}| \mathbf{u} \rangle &= \sum_{n \in \{n\}_M}\Psi_B(n, \textbf{u})\sum_{l=1}^L\mathcal{A}_{l,l+1}\\
	&\qquad\qquad \times \tr\left[\prod_{i=1}^M \left(t_3^{n_{(i+1)i}-1}t_2\right)\right]\eqncom
\end{aligned}
\label{eq:AMPSui}
\end{equation}
where $\{n\}_M$ denotes the usual set of ordered magnon positions $(n_1 < \cdots  < n_M)$ and $\Psi_B(n, \{u_i\})$ is the Bethe wave-function. Furthermore, the shorthand notation $n_{ij} \equiv n_i - n_j$ and $n_{M+1} \equiv n_1 + L$ is used throughout.

For any even $M$ and $k = 2$, one can compute directly the action of $\sum\mathcal{A}_{l,l+1}$ on the traces in \eqref{eq:AMPSui}:
\begin{multline}
 \sum_{l=1}^L\mathcal{A}_{l,l+1}\tr\prod_{i=1}^M \left(t_3^{n_{(i+1)i}-1}t_2\right)\\
 \stackrel{k=2}{=}
(-1)^{\frac{M}{2} + \sum_{i}n_i}2^{3-L}\left[ L + 2\sum_{i=1}^M\left(\delta_{n_{(i+1)i}=1} - 1\right)\right]\eqndot
\end{multline}
Using this, the rest of the computation can be carried out symbolically by brute force in Mathematica, at least for smaller
values of $M$.
This was done for $M = 2,4$ and leads to the conjecture
\begin{equation}
\langle \text{AMPS}| \mathbf{u} \rangle \stackrel{k=2}{=} \Big(4 L - \Delta^{(1)} \Big) \langle \text{MPS} | \mathbf{u} \rangle \eqncom
\label{eq:AMPSuik2allM}
\end{equation}
which was subsequently tested numerically up to and including $M = 6$ and $L = 16$.
A closed formula for $M=2$ and
any $k$ can likewise be obtained.

\paragraph{Two-loop Bethe states} 
The first loop correction to the Bethe state, i.e.\ the two-loop Bethe state, can be generated via the so-called $\Theta$-morphism \cite{Gromov:2012uv}.
To this end, we consider the Heisenberg spin chain with impurities $\theta_i$. The one-loop Bethe state can again be constructed using the algebraic Bethe ansatz approach: 
\begin{equation}
 |\theta;\mathbf{u}\rangle=\hat{B}(u_1)\dots \hat{B}(u_M)|0\rangle_L\,,
\end{equation}
where the $\hat{B}$-operator is
\begin{equation}
 \hat{B}(u)=\langle \uparrow\!|\bigotimes_{j=1}^L\left(\idm_{j,0}+\frac{i}{u-\theta_j-\frac{i}{2}}\mathbb{P}_{j,0}\right)|\!\downarrow\rangle\eqndot
\end{equation}
The two-loop eigenstate is then
\begin{equation}
 |\mathbf{u}\rangle\equiv \left(1- g^2\frac{\Delta^{(1)}}{2}\mathbb{H}_{L,1}\right)\Big\{|\theta;\mathbf{u}\rangle\Big\}_\Theta\,,
\end{equation}
where $\mathbb{H}_{j,j+1}=\idm_{j,j+1}-\mathbb{P}_{j,j+1}$ is the Heisenberg spin chain Hamiltonian density. The $\Theta$-morphism $\{\}_\Theta$ is defined via
\begin{equation}
 \Big\{f\Big\}_\Theta\equiv f+\frac{g^2}{2}\sum_{i=1}^L\bigg[\frac{\partial}{\partial\theta_i}-\frac{\partial}{\partial\theta_{i+1}}\bigg]^2f+O(g^4)\bigg|_{\theta_j\to0}\eqndot
\end{equation}
The rapidities $\{u_i\}$ have to satisfy the two-loop Bethe equations \eqref{eq:BAEquantum}. For instance, the easiest case is $M=2,k=2$, where we find for the overlap with the matrix product state
\begin{align}
\frac{\langle \text{MPS}| \mathbf{u} \rangle}{\sqrt{\langle\mathbf{u}|\mathbf{u}\rangle}} = \sqrt{\frac{L}{L-1}\frac{u^2+\frac{1}{4}}{u^2}\frac{1+g^2\frac{4}{u^2+\frac{1}{4}}}{
1+\frac{g^2}{L-1}\frac{6u^2-\frac{1}{2}}{(u^2+\frac{1}{4})^2}}}\eqndot
\end{align}
A closed expression for $M=2$ and any $k$ can similarly be derived.

\paragraph{General formula} Now that we have all the ingredients, we are ready to check if \eqref{eq:Ansatz} reproduces \eqref{eq:one-loop}. Indeed, one can analytically show that for $M=2$ both formulas agree. 
Moreover, we numerically compared \eqref{eq:Ansatz} and \eqref{eq:one-loop} for $L=8$ and $M=4$ excitations for various values of $k$ and again found perfect agreement.

\subsection{BMN vacuum at all loop orders}

A particularly simple situation arises if we consider the spin-chain vacuum, which corresponds to the 
protected operator $\tr(X^L)$.

\paragraph{} For the vacuum, there are no Bethe roots and our proposal \eqref{eq:SU2genK}  reduces to:
\begin{equation}
C_k = i^L T_{k-1}(0) =\sum_{a=\frac{1-k}{2}}^{\frac{k-1}{2}} (i g x(ia))^L\eqncom
\label{Ck_sum}
\end{equation}
i.e.\ the only contribution stems from the transfer matrix for the vacuum.  We notice, in particular, that the contribution from the flux factor trivializes. 

For even $k$ and even $L$, the one-point function formula can be readily expanded as a power series in $g$ with the result (up to order $g^{2L}$ the result is identical for odd $k$)
\begin{align}\label{eq:result}
&C_k (g)= 2 \sum^{\frac{L}{2}}_{n=0} \binom{L-n}{n}\frac{L}{L-n} \frac{B_{L-2n+1}(\tfrac{1+k}{2})}{L-2n+1}g^{2n} \\
& + g^{2L} \sum^{\infty}_{n=0} \frac{L[\Psi ^{(L+2 n-1)}(\frac{1+k}{2})-\Psi^{(L+2n-1)}(\frac{1-k}{2})]}{(-1)^n\, n!(L+n)!} g^{2n}\eqncom\nonumber
\end{align}
where $B_n$ is the Bernoulli polynomial with index $n$ and $\Psi^{(n)}$ is the polygamma function.
 We notice that the term occurring in the 
second line of~(\ref{eq:result}) only starts contributing at wrapping order. For even $k$ and odd $L$ the one-point function vanishes
as $x(u)$ is an odd function (away from the cut). At one-loop level, we find that (\ref{eq:result}) exactly agrees with \cite{Buhl-Mortensen:2016jqo}.  For odd $k$, the contribution from $a=0$  in (\ref{Ck_sum}) should be understood in the following way:
\begin{align}
\left.\left(gx(ia)\right)^L\right|_{a=0} \equiv  \left(gx(+0i)\right)^L + \left(gx(-0i)\right)^L.
\end{align}
This prescription can be motivated by the fact that it leads to the correct result for $\langle \tr X^2\rangle$ at one-loop order for odd
$k$ and in addition ensures that the one-point function vanishes  for odd $L$, also when $k$ is odd.

\paragraph{String theory} We can compare this result to a string-theory prediction in the double-scaling limit proposed in \cite{Nagasaki:2011ue}.
This limit consists in taking 
\begin{align}
 &\lambda\to\infty\eqncom && k\to \infty \eqncom && \frac{\lambda}{k^2}\,\,\text{ fixed and small,}
\end{align}
on top of the planar limit. In \cite{Nagasaki:2012re}, the one-point function of a specific
SO$(3)\times $SO(3)-invariant chiral primary was calculated by a variant of the Witten prescription, in particular implying a supergravity approximation, which is justified here due to the assumption of
$\lambda\rightarrow \infty$. As explained in~\cite{Buhl-Mortensen:2016jqo}, the result of this computation can be turned into
a prediction for the one-point function we are considering divided by its tree-level value. 

The prediction from string theory reads
\begin{align}
\left.\frac{C_k(g)}{C_k(0)}\right|_{st}&=\frac{{\mathrm\Gamma}(L+\frac{1}{2})}{\kappa^{L+1} \sqrt{\pi}{\mathrm\Gamma}(L)}\left[\kappa^2+1\right]^{\frac{3}{2}}  \\
& \phaneq\times \int_{-\arctan\kappa}^{\frac{\pi}{2}} \!\!\!\!\!\!\!\! \de \theta  \cos^{2L-1}\theta \left(\kappa+\tan\theta\right)^{L-2}\eqndot \nonumber 
\end{align}
The leading two terms of the integral above in the large $\kappa=\frac{\pi k}{\sqrt{\lambda}}$ expansion were already given in \cite{Nagasaki:2012re} and we can even evaluate the integral exactly to get
\begin{equation}
\left.\frac{C_k(g)}{C_k(0)}\right|_{st}=
\frac{ \left(\kappa+\sqrt{\kappa^2+1}\right)^L \left(L \sqrt{\kappa^2+1} -\kappa\right)}{2^L(L-1)\kappa^{L+1}}\eqndot
\end{equation}
\paragraph{Comparison} Let us ignore the second line of \eqref{eq:result} which as mentioned above only starts contributing at
wrapping order. In the large-$k$ limit, we have  $B_n(\frac{1+k}{2})\rightarrow (\frac{k}{2})^n$ and the first line of \eqref{eq:result}
organizes itself as a power series in $\left(\frac{g}{k}\right)^2$
\begin{align}
\left.\frac{C_k(g)}{C_k(0)}\right|_{gt} {\longrightarrow} \,\,\,\,\,&1+ \sum^{\frac{L}{2}}_{n=1} \binom{L-n}{n-1}\frac{L}{n}\frac{L+1}{L-n}\left( \frac{2g}{k} \right)^{2n} \nonumber\\
&+{\cal O}(g^{2L}), \hspace{0.5cm}\mbox{as $k\rightarrow \infty$}
\eqndot 
\end{align}
Remarkably, this agrees  with the string-theory prediction up to wrapping order after identifying $\kappa= \frac{k}{4g}=\frac{\pi k}{\sqrt{\lambda}}$.
The terms in the second line of \eqref{eq:result} have a scaling behaviour in $k$ which violates  the double-scaling limit. It is tempting to attribute
these terms to wrapping interactions.

\subsection{Flux factor}

The flux factor in our proposal \eqref{eq:Ansatz} has no counterpart at tree level and depends on the anomalous scaling dimension $\Delta-L$ such that it vanishes for protected operators.

At one-loop order, the corresponding contribution in \eqref{eq:one-loop} has been calculated in \cite{Buhl-Mortensen:2016jqo}. It is the finite part of the UV-divergent integral whose UV divergence is subtracted by the renormalization constant and yields the one-loop scaling dimension $\Delta^{(1)}$.
Since UV divergences exponentiate,  it is possible that the flux factor exponentiates as well,  and the following form of the higher loop flux factor seems natural
\begin{equation}
\label{eq: flux exponentiation}
 \mathbb{F}_k=2^{L-\Delta}\exp\Big[(\Delta-L)(\digamma({\textstyle\frac{k+1}{2}})+\gamma_E)\Big]\eqndot
\end{equation}
A direct field-theoretic check of \eqref{eq: flux exponentiation} at two-loop order would clearly be desirable, though very demanding.

An independent consequence of the flux factor is that it leads to a breakdown of the double-scaling limit for non-protected operators starting already at one-loop order.
As an example, let us consider the Konishi operator, which has $L=4,M=2$ and $u_1=-u_2=\frac{1}{2\sqrt{3}}+O(g^2)$.
Its one-loop one-point function can be explicitly worked out to be 
\begin{align}
C_k = \frac{k(k^2-1)}{12\sqrt{3}}\Big(&1 + 12g^2\left[\Psi(\tfrac{k+1}{2}) + \gamma_E - \log 2 + \tfrac{5}{6} \right]
\Big)\,,
\end{align} 
where we used that $\Delta^{(1)} = 12$.
Since $\Psi(\tfrac{k+1}{2})\sim \log k$ for large $k$, the perturbative expansion in the double-scaling limit does not arrange itself in powers of $\frac{\lambda}{k^2}$. 

\section{Conclusions \& Outlook}

We have argued that the recently derived, integrability-based formula for  tree-level one-point functions in 
the SU$(2)$ sector of a specific defect version of ${\cal N}=4$ SYM theory points towards a natural higher-loop generalization.  The generalization is based on an idea which worked successfully for the spectral problem of ${\cal N}=4$ SYM theory, and which consists of introducing the coupling constant via a Zhukovski transformation of the Bethe roots characterizing the conformal operators.
More precisely, the Zhukovski variables should replace the Bethe roots both in the Bethe equations and the transfer matrix of the system
and the Bethe equations should be equipped with the usual phase factor of ${\cal N}=4$ SYM theory.
Furthermore, in the present case an additional flux factor contributing to the higher-loop one-point function formula is needed.

We have performed a number of non-trivial  consistency checks of the generalized one-point function formula and these have come out positive.
First, we have compared the higher-loop one-point function formula to an honest field-theory calculation of the one-loop one-point function of non-protected operators in the SU(2) sector. This calculation is technically demanding, involving the evaluation of the overlap of an uncorrected Bethe eigenstate and a so-called amputated matrix product state as well as the overlap between a loop-corrected Bethe eigenstate and an uncorrected matrix product state~\cite{Buhl-Mortensen:2016jqo}.  Results can be obtained analytically 
for BMN operators with two excitations, whereas for more complicated operators one has to resort to numerical
computations. For all cases tested, the field-theory computation agreed with the proposed higher-loop
formula. As a second test, we have carried out an analysis of the higher-loop formula when applied
to the BMN vacuum state $\tr(X^L)$. For this state, the one-point function consists of two contributions, one which comes into play
only at wrapping order and one for which it is possible to impose  the double-scaling limit,
proposed in \cite{Nagasaki:2011ue}, and obtain a power
series expansion in the double-scaling parameter.
 This power series expansion can be compared to a similar expansion 
obtained by a string-theory analysis using a supergravity approximation and  agreement is found up to wrapping order for
any length, $L$, of the BMN vacuum state.
These two consistency checks constitute a strong indication
that we are on the right track when trying to move towards higher loop orders. 

The flux factor we propose depends on the anomalous dimension of the operator considered and leads to a breakdown of the double-scaling limit in the case of non-protected operators starting already at one-loop order. 
While the exponentiation of the flux factor is certainly natural from the one-loop point of view, an explicit field-theoretic check at two-loop order is clearly required.

The presented higher-loop one-point function formula is expected to be only an asymptotic formula in the sense
that we expect there to be further corrections from  wrapping interactions as it was the case for ${\cal N}=4$ SYM theory~\cite{Beisert:2004hm,Ambjorn:2005wa}.  
It would be very interesting to investigate the possible wrapping corrections in the present defect CFT or to study the theory using the thermodynamical Bethe ansatz approach to clarify whether the
second line of (\ref{eq:result}) can indeed be understood as wrapping terms.

It would likewise be interesting to investigate whether the integrability approach can be used to infer some properties of the higher-loop contributions to other observables in the present defect CFT such as Wilson 
loops~\cite{Nagasaki:2011ue, deLeeuw:2016vgp,Aguilera-Damia:2016bqv}
or  less studied  objects such as two-point functions of operators 
of unequal conformal dimension~\cite{deLeeuw:2017dkd}.

\begin{acknowledgments}
\paragraph{Acknowledgements.}
I.B.-M.,  M.d.L., C.K.\  and M.W.\ were  supported  in  part  by  FNU  through grant number DFF-4002-00037. 
\end{acknowledgments}

\bibliography{ABAletter}

\end{document}